\documentclass[twocolumn,showpacs,preprintnumbers,amsmath,amssymb]{revtex4}
\usepackage{graphicx}
\usepackage{dcolumn}
\usepackage{bm}

\begin{document}

\title{Temperature of a Compressed Bubble with Application to Sonoluminescence}

\author{S. Sivasubramanian}
\author{A. Widom}
\author{Y. N. Srivastava}
\affiliation{Physics Department, Northeastern University, Boston MA 02115}

\date{\today}

\begin{abstract}
The rise in temperature from the adiabatic compression 
of a bubble is computed in thermodynamic mean field 
(van der Waals) theory. It is shown that the temperature 
rise is higher for the noble gas atoms than for more 
complex gas molecules. The adiabatic temperature rise is 
shown to be sufficient for producing sonoluminescence via 
the excited electronic states of the atoms. 
\end{abstract}

\maketitle

\section{Introduction}

There has been considerable interest in the dynamics of single bubble 
sonoluminescence\cite{Survey1,Survey2,Survey3,Survey4}. 
It is found that noble gases (within the bubble) greatly 
enhance the emitted light intensity when compared to 
bubbles filled with other gases 
\cite{RaregasWater,WaterVapour,Airwater,Inertgasimp,InertGas1,
InertGas2,InertGas3}. The theoretical interpretation of this 
experimental fact will be discussed in the work which follows.
 
The fluid mechanics of bubble motion\cite{Bubble1,Bubble2,Bubble3} 
is thought to play a central role in the dynamics of sonoluminescence.
The starting point of theoretical bubble dynamics has been 
the Rayleigh-Plesset equation \cite{Bubble4,Bubble5,Bubble6}. 
An important feature of the experimental bubble fluid 
dynamics is the exceedingly high 
temperatures\cite{Hydrocarbon,Hydrogen,Infrared} 
that can be achieved by merely decreasing the radius of a 
bubble. This is especially true when noble gas atoms are contained 
within the bubble 
\cite{NobleGas1,NobleGas2,NobleGas3,NobleGas4,Temperature1,Temperature2}.  
Although the dynamics of bubble 
compression appears to be intimately connected to the high noise 
temperature of emitted light pulses, the process of light 
emission has not yet been firmly established 
\cite{Model1,Model2,Model3,Model4,Model5}.
 
Our purpose is to calculate the maximum 
temperature of the contents inside the bubble on the basis of
fluid mechanics\cite{Hot1,Hot2,Hot3} using a mean field 
(van der Waals) thermodynamic equation of state for adiabatic 
compression. We find that the resulting temperatures are 
sufficiently high to produce light from sonoluminescence 
if not from mere incandescence. We also estimate the concentration 
of atoms with excited electronic states which arise from an adiabatic 
compression of the fluid within the bubble.

In Sec.II, we derive the thermodynamic identities required to 
compute the temperature rise during adiabatic compression of a bubble. 
The mean field (van der Waals) thermodynamic equation of state 
is discussed in Sec.III wherein numerical results are presented 
for the case of argon. Arguments are presented which explain why 
noble gas atoms are particularly suited for achieving high temperatures 
for small bubble radii. In Sec.IV, a Lagrangian formulation of 
the bubble dynamics will be employed to discuss in more detail 
the role of adiabatic processes. The internal 
chemistry\cite{ChemicalKinetics} of the bubble 
will be discussed in Sec.V, while the physical   
aspects of the chemical reaction will be further explored in the 
concluding Sec.VI.

\section{Thermodynamics}

If \begin{math} \epsilon (s,v) \end{math} is the energy per 
particle of a fluid {\it inside the bubble} as a function of 
volume per particle \begin{math}v \end{math} and entropy per 
particle \begin{math}s \end{math}, then 
\begin{equation}
d{\epsilon}=Tds-pdv,
\label{Thermo1}
\end{equation}
where \begin{math}T \end{math} and \begin{math}p \end{math} 
denote (respectively) the temperature and pressure of the fluid. 
Alternatively, one may employ the free energy per particle 
\begin{equation}
f(T,v)=\inf_{\epsilon}\{\epsilon-Ts(\epsilon,v)\},
\label{Thermo2}
\end{equation} 
which obeys 
\begin{equation}
df=-sdT-pdv.
\label{Thermo3}
\end{equation}

Condensed matter often (but not always) exhibits a lowering 
of the temperature when adiabatically expanded and a temperature 
rise when adiabatically compressed. The thermal coefficient 
describing such a process may defined as  
\begin{equation}
\eta=-\left(\frac{v}{T} \right) 
\left(\frac{\partial T}{\partial v} \right)_s.   
\label{Thermo4}
\end{equation}
From Eq.(\ref{Thermo1}) it follows that
\begin{equation}
-\left(\frac{\partial T}{\partial v} \right)_s
=\left(\frac{\partial p}{\partial s} \right)_v
=\left(\frac{\partial p}{\partial T} \right)_v
 \left(\frac{\partial T}{\partial s} \right)_v.         
\label{Thermo5}
\end{equation}
Eqs.(\ref{Thermo4}) and (\ref{Thermo5}) imply
\begin{equation}
\eta=\frac{v}{c_v}
\left(\frac{\partial p}{\partial T} \right)_v,
\label{Thermo6}
\end{equation}
where 
\begin{equation}
c_v=T\left(\frac{\partial s}{\partial T} \right)_v
\label{Thermo7}
\end{equation}
is the specific heat at constant volume. Similarly
\begin{equation}
c_p=T\left(\frac{\partial s}{\partial T} \right)_p
\label{Thermo8}
\end{equation}
is the specific heat at constant pressure which obeys 
\begin{equation}
c_p=c_v+T\left(\frac{\partial v}{\partial T} \right)_p
\left(\frac{\partial p}{\partial T} \right)_v.
\label{Thermo9}
\end{equation}
On the one hand, we have from Eqs.(\ref{Thermo4}), (\ref{Thermo5})  
and (\ref{Thermo7}) that 
\begin{equation}
-\left(\frac{\partial T}{\partial v} \right)_s=
\frac{T}{c_v}\left(\frac{\partial p}{\partial T} \right)_v .
\label{Thermo10}
\end{equation}
On the other hand, with 
\begin{equation}
\gamma =\left(\frac{c_p}{c_v}\right)
\ \ {\rm and}\ \ \beta=
\frac{1}{v}\left(\frac{\partial v}{\partial T}\right)_p 
\label{Thermo11}
\end{equation}
Eqs.(\ref{Thermo9}), (\ref{Thermo10}) and (\ref{Thermo11}) 
imply 
\begin{equation}
-\left(\frac{\partial T}{\partial v} \right)_s=
\left(\frac{\gamma -1}{v\beta }\right) .
\label{Thermo12}
\end{equation}
The temperature rise corresponding to an adiabatic compression 
is then calculated from 
\begin{equation}
\ln\left(\frac{T_f}{T_i}\right)=-\int_{v_f}^{v_i}
\frac{1}{T}\left(\frac{\partial T}{\partial v} \right)_s dv, 
\label{Thermo13}
\end{equation}
where \begin{math} (\partial T/\partial v)_s  \end{math} may be 
calculated from either Eq.(\ref{Thermo10}) or Eq.(\ref{Thermo12}).

\begin{figure}[bp]
\scalebox {0.5}{\includegraphics{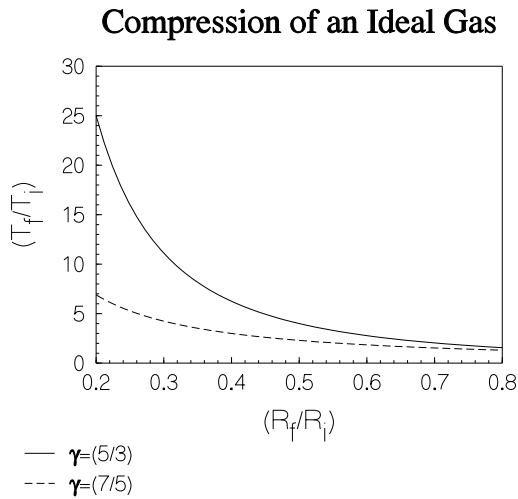}}
\caption{For the adiabatic compression of an ideal gas in a bubble 
from radius $R_i\to R_f<R_i$, the temperature rise is described by 
$T_i\to T_f>T_i$. For the case of noble gas atoms ($\gamma =5/3$) 
the temperature rise is much more steep than the case of tumbling 
diatomic molecules ($\gamma =7/5$).}
\label{fig1}
\end{figure}

For the case of an ideal gas with \begin{math} P_0 v=k_BT \end{math}, 
\begin{math} \gamma=const \end{math} and
\begin{math} T\beta_0 =1 \end{math}, Eqs.(\ref{Thermo12}) 
and (\ref{Thermo13}) imply 
\begin{equation}
\left(\frac{T_f}{T_i}\right)_{(ideal\ gas)}
=\left(\frac{v_i}{v_f}\right)^{(\gamma -1)}=
\left(\frac{R_i}{R_f}\right)^{3(\gamma -1)},
\label{Thermo14}
\end{equation}
where \begin{math} N \end{math} particles in a bubble of radius 
\begin{math} R \end{math} are described by the volume 
\begin{equation}
V=Nv=\left(\frac{4\pi R^3}{3}\right).
\label{Thermo15}
\end{equation}
The adiabatic compression of an ideal gas is plotted in
FIG.\ref{fig1}. For noble gas atoms (\begin{math} \gamma=5/3 \end{math}) 
the rise in temperature is substantial \cite{Idealgas}. 
If the radius drops by a factor 
of five, the ideal gas can exhibit a temperature rise from (say) 
room temperature to a temperature 
\begin{math} T\sim 7\times 10^3\ K \end{math} 
\cite{Predicttemp1,Predicttemp2,Predicttemp3,Predicttemp4,Predicttemp5,
Predicttemp6}
which is approximately the surface temperature of the sun. 
For an ideal gas of tumbling diatomic molecules 
(\begin{math} \gamma=7/5 \end{math}) the rise in temperature upon 
adiabatic compression is much less steep. 

The general situation follows from the thermodynamic 
Eqs.(\ref{Thermo10}) and (\ref{Thermo13}). For a complicated molecule, 
the internal degrees of freedom of the constituent molecules give rise 
to a high specific heat \begin{math} c_v  \end{math}. 
The adiabatic compression of a fluid with a high heat capacity 
yields an only moderate rise in temperature. On the other hand, 
the noble gas atoms have at most internal electronic degrees of 
freedom and these lie dormant for reasonably low 
temperatures. Thus, noble gas atom fluids have a low heat capacity and 
a resulting high temperature rise on an adiabatic compression. These 
qualitative considerations give more than just a little insight into why 
noble gas atoms (in particular Ar atoms) have played such an important 
role in laboratory experiments on sonoluminescence. 

The sonoluminescent bubble can be so compressed that the fluid
contained within the bubble is far from ideal. However, the 
above thermodynamic considerations are general and can be 
applied to fluids which are far from being ideal gases. 
In the mean field (van der Waals) theory of 
argon fluids, the equations of state can be worked out in detail.

\section{van der Waals Theory} 

The free energy per particle of an ideal gas of noble gas atoms 
may be written as 
\begin{equation}
f_0 (T,v)=-k_BTln\left(\frac{v}{\lambda_T^3}\right),
\label{vdw1}
\end{equation} 
where the thermal wavelength of an atom of mass 
\begin{math} m \end{math} is given by 
\begin{equation}
\lambda_T =\sqrt{\frac{2\pi \hbar^2}{mk_BT}}\ .
\label{vdw2}
\end{equation} 
In mean field theory, the hard core of the two atom potential 
is included by the ``free volume'' replacement 
\begin{math} v\to v-b \end{math} while the attractive part 
of the potential is simulated by subtracting (from the free energy) 
an attractive mean potential term \begin{math} (a/v) \end{math} 
proportional to the fluid density. Thus, the ``mean field'' theory 
modification to the ideal gas Eq.(\ref{vdw1}) is given by 
\begin{equation}
\bar{f} (T,v)=-k_BTln\left(\frac{v-b}{\lambda_T^3}\right)
-\left(\frac{a}{v}\right).
\label{vdw3}
\end{equation}
The mean field theory pressure 
\begin{math} \bar{p}=-(\partial \bar{f}/\partial v)_T  \end{math} 
which follows from Eq.(\ref{vdw3}) is that of van der Waals 
\begin{equation}
\bar{p}=\left(\frac{k_BT}{v-b}\right)-\left(\frac{a}{v^2}\right),
\label{vdw4}
\end{equation}
while the specific heat at constant volume follows from Eqs.(\ref{vdw2}) 
and (\ref{vdw3}) as 
\begin{equation}
\bar{c}_v=T\left(\frac{\partial \bar{s}}{\partial T}\right)_v
=-T\left(\frac{\partial^2 \bar{f}}{\partial T^2}\right)_v
=\left(\frac{3k_B}{2}\right).
\label{vdw5}
\end{equation} 
Eqs.(\ref{Thermo10}), (\ref{vdw4}) and (\ref{vdw5}) imply that 
\begin{equation}
-\left(\frac{\partial T}{\partial v} \right)_s
=\left(\frac{2T}{3(v-b)}\right)\ \ {\rm (van\ der\ Waals)}.
\label{vdw6}
\end{equation} 

\begin{figure}[bp]
\scalebox {0.5}{\includegraphics{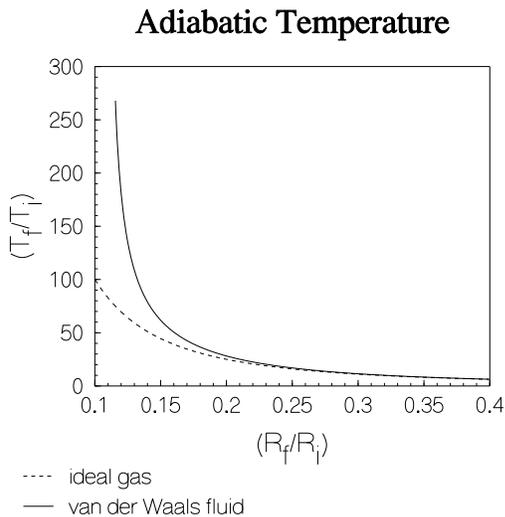}}
\caption{In the adiabatic compression of a fluid
within a spherical bubble from radius $R_i$ to radius $R_f$, 
the temperature rise from $T_i$ to $T_f$ is plotted. The temperature 
rise from room temperature when interactions are included (as in 
the van der Waals theory) is large compared with the ideal gas case but 
only if the compression is substantial.}
\label{fig2}
\end{figure}

Finally, Eqs.(\ref{Thermo13}) and (\ref{vdw6}) yield the mean field 
theory adiabatic temperature rise in a noble gas atom fluid upon 
adiabatic compression; It is 
\begin{eqnarray}
\left(\frac{T_f}{T_i}\right)_{(van\ der\ Waals)}&=&
\left(\frac{v_i-b}{v_f-b}\right)^{2/3} \nonumber \\ 
&=& \left(\frac{1-y^3}{(R_f/R_i)^3-y^3}\right)^{2/3},
\label{vdw7}
\end{eqnarray}
where Eq.(\ref{Thermo15}) has been invoked and the initial volume 
per particle \begin{math} v_i \end{math} determines  
\begin{equation}
y=(b/v_i)^{1/3}=(\xi /\Lambda ).
\label{vdw8}
\end{equation}
In Eq.(\ref{vdw8}), \begin{math} b=\xi^3 \end{math} and 
\begin{math} v_i=\Lambda^3 \end{math}. For argon \cite{Thermodynamic}, 
\begin{math} \xi \approx 3.7664\ \AA \end{math} and for an 
initial state at room temperature and pressure  
\begin{math} \Lambda \approx 34.356\ \AA \end{math}.
Thus \begin{math} y(Ar) \approx 0.1096 \end{math}. 
A plot of Eq.(\ref{vdw7}) is shown in FIG.\ref{fig2}.

If the final bubble radius \begin{math} R_f  \end{math} is only 
(say) \begin{math} \approx 10\% \end{math} of the initial radius 
\begin{math} R_i \end{math} \cite{Radius1, Radius2, Radius3}, 
then in the van der Waals model 
the temperature rise can be enormous (say) from room temperature 
to a final temperature of  
\begin{math} \approx 7.5\times 10^4\ K \end{math} at which point 
the fluid would appear to be incandescent. 

\section{Lagrangian Dynamics}

For the purpose of better understanding the role of adiabatic processes 
in bubble dynamics, we here consider such dynamics from an analytical 
mechanics viewpoint. The fluid mechanics of a spherical bubble can 
be described by a Lagrangian of the form 
\begin{equation}
L(\dot{R},R)=\frac{1}{2}{\cal M}(R)\dot{R}^2-{\cal U}(R)-
{\cal U}_{ext}(R),
\label{L1}
\end{equation} 
where the external driving potential for the bubble 
dynamics is denoted by \begin{math} {\cal U}_{ext} \end{math} 
and a Rayleigh dissipation function \cite{dissipation} 
\begin{math} {\cal F} \end{math} 
is employed having the form
\begin{equation}
2{\cal F}(\dot{R},R)=\Gamma (R)\dot{R}^2.
\label{L2}
\end{equation}
The dynamical equation of motion reads 
\begin{equation}
\frac{d}{dt}\left(\frac{\partial L}{\partial \dot{R}}\right)=
\left(\frac{\partial L}{\partial R}\right)-
\left(\frac{\partial {\cal F}}{\partial \dot{R}}\right).
\label{L3}
\end{equation}
The various terms contributing to Eqs.(\ref{L1}) and (\ref{L2}) 
will now be explored.

During an expansion or compression of a spherical bubble, 
the fluid velocity field outside the bubble is  
\begin{equation}
{\bf v}=
\left(\frac{\bf r}{|{\bf r}|^3}\right)R^2\dot{R}
\ \ \ {\rm for}\ \ \ (|{\bf r}|>R)
\label{L4}
\end{equation}
while the fluid velocity field inside the bubble is 
\begin{equation}
{\bf v}=
\left(\frac{\dot{R}}{R}\right){\bf r}
\ \ \ {\rm for}\ \ \ (|{\bf r}|<R).
\label{L5}
\end{equation}
The (presumed) incompressible flow outside the bubble 
requires a kinetic energy
\begin{eqnarray}
K_{out}&=&\frac{\rho}{2}\int_{|{\bf r}|>R}|{\bf v}|^2d^3{\bf r}
\nonumber \\
&=&\frac{\rho}{2}\int_R^\infty \left|\frac{R^2\dot{R}}{r^2}\right|^2 
(4\pi r^2dr)\nonumber \\ 
&=&2\pi \rho R^3\dot{R}^2, 
\label{L6}
\end{eqnarray}
while the flow of the fluid inside the bubble requires a 
kinetic energy
\begin{eqnarray}
K_{in}&=&\frac{M}{2(4\pi R^3/3)}
\int_{|{\bf r}|<R}|{\bf v}|^2d^3{\bf r}
\nonumber \\
&=&\frac{3M}{8\pi R^3}
\int_0^R \left|\frac{r\dot{R}}{R}\right|^2 
(4\pi r^2dr)\nonumber \\ 
&=&(3/10)M\dot{R}^2, 
\label{L7}
\end{eqnarray}
where \begin{math} M \end{math} denotes the mass of fluid 
(presumed) conserved within the bubble.
From Eqs.(\ref{L6}) and (\ref{L7}) one finds a total fluid flow 
kinetic energy of 
\begin{equation}
K=K_{in}+K_{out}=\frac{1}{2}{\cal M}(R)\dot{R}^2 
\label{L8}
\end{equation}
for which 
\begin{equation}
{\cal M}(R)=\frac{3}{5}M+4\pi \rho R^3.
\label{L9}
\end{equation}
Eq.(\ref{L9}) establishes the effective mass 
\begin{math} {\cal M}(R) \end{math} to be used in the 
Lagrangian Eq.(\ref{L1}).

For the potential terms in the Lagrangian Eq.(\ref{L1}), one may 
write a sum of the internal energy of the fluid within 
the bubble and a term proportional to the surface area 
\begin{equation}
{\cal U}(R)=N\epsilon \left(s=\frac{S}{N},v=\frac{4\pi R^3}{3N}\right)
+4\pi \sigma R^2.
\label{L10}
\end{equation}
The external potential may be written in terms of the 
fluid pressure \begin{math} P \end{math} exerted on the bubble 
by the external fluid; i.e.  
\begin{equation}
{\cal U}_{ext}(R)=\left(\frac{4\pi R^3P}{3}\right).
\label{L11}
\end{equation}
Eqs.(\ref{L9}), (\ref{L10}) and (\ref{L11}) completely specify 
the Lagrangian in Eq.(\ref{L1}).

To compute the dissipation function, we note for the 
incompressible flow (with viscosity 
\begin{math} \eta  \end{math}) \cite{viscous} outside the bubble 
\begin{eqnarray}
2{\cal F}_{out}&=&\frac{\eta }{2}
\int_{|{\bf r}|>R}(\partial_i v_j+\partial_j v_i)^2d^3{\bf r}
\nonumber \\
&=& 16\pi \eta R\dot{R}^2, 
\label{L12}
\end{eqnarray}
where Eq.(\ref{L4}) has been invoked. For the flow inside the 
bubble, only the bulk viscosity \begin{math} \zeta  \end{math} 
\cite{viscous} enters into the computation, i.e. 
\begin{eqnarray}
2{\cal F}_{in}&=& \zeta \int_{|{\bf r}|<R}(div{\bf v})^2d^3{\bf r}
\nonumber \\
&=& 12\pi \zeta R\dot{R}^2, 
\label{L13}
\end{eqnarray}
where Eq.(\ref{L5}) has been invoked. From Eqs.(\ref{L12}) and 
(\ref{L13}) one finds a total dissipation function  
\begin{equation}
2{\cal F}=2{\cal F}_{in}+2{\cal F}_{out}=\Gamma (R)\dot{R}^2
\label{L14}
\end{equation}
where 
\begin{equation}
\Gamma (R)=16\pi\left(\eta +\frac{3}{4}\zeta \right)R.
\label{L15}
\end{equation}
Eq.(\ref{L15}) establishes the friction coefficient
\begin{math} \Gamma (R) \end{math} to be used in
the dissipation function Eq.(\ref{L2}).

Employing Eqs.(\ref{Thermo1}), (\ref{L1}-\ref{L3}), 
(\ref{L9}-\ref{L11}) and (\ref{L15}) we find the 
following equation of motion: 
\begin{eqnarray}
(1+\alpha_1)R\ddot{R}+\left(\frac{3\dot{R}^2}{2}\right)+
4(1+\alpha_2)\left(\frac{\nu \dot{R}}{R}\right)
\nonumber \\ 
=-\left(\frac{P+(2\sigma /R)-p}{\rho }\right), 
\label{L16}
\end{eqnarray}
where 
\begin{equation}
\nu =(\eta /\rho)
\label{L17}
\end{equation}
is the kinematic viscosity of the liquid outside the bubble, 
and \begin{math} \alpha_i  \end{math} for 
\begin{math} i=1,2 \end{math} in Eq.({\ref{L16}}) 
represent the corrections to the usual  
Rayleigh-Plesset equation \cite{Bubble4,Bubble5,Bubble6} 
due to motions within the bubble. 
In detail, 
\begin{equation}
\alpha_1=\left(\frac{3M}{20\pi \rho R^3}\right)
\ \ \ {\rm and}\ \ \ \alpha_2=\left(\frac{3\zeta }{4\eta }\right).
\end{equation}
In the usual approximation, 
\begin{math} \alpha_2\approx 0 \end{math}, 
all of the dissipative heating is from outside the bubble, 
and the  compressions and/or expansions of the bubble are 
strictly adiabatic. For the internal heating of fluid inside 
the bubble, \begin{math} \alpha_2>0 \end{math}, the bulk
viscosity \begin{math} \zeta \end{math} of the fluid inside 
the bubble will give rise to additional internal heating. 

\section{Electronic Excitation of Atoms}

Let us now further consider the case in which the fluid inside 
the bubble is made up of argon atoms. We wish to consider the physical 
chemistry of a simple reaction in which the electron states of  
argon atoms are excited; i.e.
\begin{equation}
Ar \leftrightarrows Ar^*, 
\label{C1}
\end{equation}
where \begin{math} Ar \end{math} represents an atom in the 
ground electronic state while \begin{math} Ar^* \end{math} 
represents an atom the atom in an excited electronic state. 
The concentration of atoms in an excited electronic state 
is defined as 
\begin{equation}
x=\left(\frac{N_{Ar^*}}{N_{Ar}+N_{Ar^*}}\right). 
\label{C2}
\end{equation}

The second law of thermodynamics dictates that 
chemical equilibrium arises when the energy reaches a minimum 
for constant entropy; i.e.
\begin{equation}
\epsilon (s,v)=\inf_{0\le x\le 1}\tilde{\epsilon }(s,v,x).
\label{C3}
\end{equation}
When the fluid in the bubble is out of chemical equilibrium, 
Eq.(\ref{Thermo1}) is generalized to 
\begin{equation}
d\tilde{\epsilon }=Tds-pdv+\tilde{\mu }dx,
\label{C4}
\end{equation}
where 
\begin{equation}
\tilde{\mu }=\mu_{Ar^*}-\mu_{Ar}.
\label{C5}
\end{equation}
If \begin{math} x<<1 \end{math}, then it is sufficiently accurate 
to use the conventional dilute solution theory wherein 
\begin{equation}
\tilde{\mu }=\phi +k_BT\ln x
\label{C6}
\end{equation} 
and the activation free energy \begin{math} \phi \end{math} 
is independent of \begin{math} x \end{math}. From Eqs.(\ref{C3}) 
and (\ref{C4}), it follows for chemical equilibrium that   
\begin{math} \tilde{\mu }=0 \end{math}. Eq.(\ref{C6}) 
then yields the thermal activation formula for concentration  
\begin{equation}
x=e^{-\phi /k_BT}\ \ \ \ {\rm (chemical\ equilibrium)}.
\label{C7}
\end{equation}

\begin{figure}[bp]
\scalebox {0.5}{\includegraphics{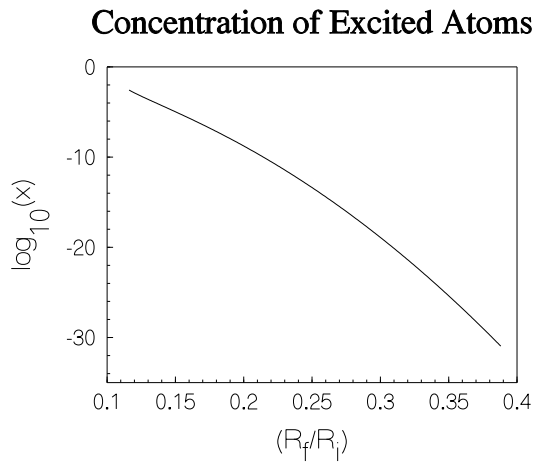}}
\caption{As bubble containing pure argon is adiabatically compressed 
from an initial radius $R_i$ at room temperature and pressure to a final 
radius $R_f<R_i$, the concentration $x$ of electronically excited 
argon atoms sharply increases.}
\label{fig3}
\end{figure}

In thermodynamic mean field theory, the activation free energy 
may be written in terms of the energy states of a free 
argon atom as 
\begin{equation}
\bar{\phi }=-k_BT\ln 
\left(
\frac{\sum_{E>E_0}w_E e^{-E/k_BT}}{\sum_E w_E e^{-E/k_BT}}
\right), 
\label{C8}
\end{equation}
where \begin{math} w_E \end{math} is the degeneracy of the 
electronic energy level \begin{math} E \end{math} 
and \begin{math} E_0 \end{math} is the electronic ground state 
energy. For argon it is known\cite{excitedstates} that the first 
two electronic excited states are given by
\begin{eqnarray}
E\Big(3p^5\big(^2P_{3/2}\big)4s,J=1 \Big)-E_0
&\approx &11.63\ eV \nonumber \\
E\Big(3p^5\big(^2P_{1/2}\big)4s,J=1 \Big)-E_0
&\approx &11.82\ eV .
\label{C9}
\end{eqnarray}
Including the above electronic states in the computation of 
the activation free energy in Eq.(\ref{C8}) and employing 
Eqs.(\ref{C7}) and (\ref{vdw7}) we plot the concentration of 
excited states as a function of bubble radius in FIG.\ref{fig3}.

\section{Conclusions}

It has been shown that the adiabatic compression of noble gas atoms 
gives rise to surprisingly high temperatures. The statistical 
thermodynamics of the processes may viewed as follows: (i) The entropy 
of the fluid inside the bubble is proportional to the logarithm of the 
number of states. (ii) When the bubble is compressed, due to the hard core 
part of the inter-atomic potential, there is an excluded spatial volume. 
The compression then eliminates much available spatial volume 
part of phase space which would in principle lower the entropy. 
(iii) To keep the entropy constant during 
the adiabatic compression, the momentum part of phase space must expand 
as the position part of phase space contracts. Thus there will exist 
an increased kinetic energy. A high mean kinetic energy (in classical 
statistical thermodynamics) is equivalent to a high temperature. 
This explains why an adiabatic bubble compression can lead to such 
a high temperature. 
  
That noble gas atoms achieve even higher adiabatic compression temperatures 
than the temperatures which would be achieved with more complex molecular 
substances has also been explained. The compression lowers the spatial 
parts of phase space so other parts of phase space must be opened up to 
keep the entropy constant in an adiabatic process. For a fluid of 
noble gas atoms, momentum phase space is opened up giving rise to 
high kinetic energy and high temperature. If the fluid inside 
the bubble contains molecules with structure, then other parts 
of phase space can open up to make the lost 
spatial part of phase space. These include internal tumbling motions of 
molecules as well as internal vibrational excitations. Not all of the 
required new regions of phase space involve a rise in center of mass kinetic 
energy of the molecule which is equivalent to a rise in temperature.
Thus, a fluid composed of complex molecules will not exhibit as high 
a compression induced temperature rise as will a fluid 
made up of noble gas atoms. From a mathematical point of 
view, the specific heat 
\begin{math} c_v  \end{math} in Eq.(\ref{Thermo7}) describes how much phase 
space is opened up by a given temperature rise. Each of the internal 
degrees of freedom in a complex molecule raises 
\begin{math} c_v  \end{math}. The higher the specific heat, the lower 
the temperature rise in Eqs.(\ref{Thermo10}) and (\ref{Thermo13}). 

The bubble dynamics has been described through a Lagrangian,
including the Rayleigh dissipation function yielding the usual 
results apart from small correction terms arising from fluid 
flow within the bubble. Since the energy for fixed entropy 
enter into the Lagrangian as potential energy terms, the adiabatic 
compression holds to a high degree of accuracy. 

The concentration of electronically excited atoms at compression 
induced high temperatures can be approximately estimated in mean field 
theory. One uses an electronic partition function which is strictly valid 
only for ideal gases. When the fluid is compressed, there will in reality 
be a shift in electronic energy. The excited electron shells will be 
somewhat delocalized to neighboring atoms. In fact, one might well have 
coherent electron states in such a system giving rise to superradiance 
during the light emission pulse. However, the situation regarding the 
coherence of sonoluminescent optical pulses is still unclear and 
requires further elucidation.

\end{document}